%% file: main.tex
\documentclass[conference]{IEEEtran}

\usepackage{ifthen}
\newboolean{blind}   
\setboolean{blind}{false}

\IEEEoverridecommandlockouts
\usepackage[letterpaper, left=0.625in, right=0.625in, bottom=1in, top=0.71in]{geometry}

\usepackage{cite}
\usepackage{adjustbox}
\usepackage{amsmath,amssymb,amsfonts}
\usepackage[ruled]{algorithm2e}
\usepackage{graphicx}
\usepackage{textcomp}
\usepackage{cleveref}
\usepackage{bm}
\usepackage{comment}
\usepackage{stfloats}
\usepackage{pgfplots}
\usepackage{booktabs}
\usepackage{multirow}
\usepackage{verbatim}
\usepackage{xcolor}

\usepgfplotslibrary{colormaps}

\usepackage{siunitx}
\DeclareSIUnit\operations{Op}
\DeclareSIUnit\BL{\giga\bit\kilo\meter\per\second}
\DeclareSIUnit\FL{\decibel\per\kilo\meter}
\DeclareSIUnit\CD{\pico\second\per\nano\meter\per\kilo\meter}
\DeclareSIUnit\baud{Bd}
\DeclareSIUnit{\nothing}{\relax}

\usepackage{mleftright}
\usepackage{pgfplots}
\usepackage{ wasysym }

\usetikzlibrary{backgrounds}

\input{makros_own}

\usepackage[nohyperlinks,nolist]{acronym}
\input{acros.tex}

\usepackage{tikz}
\usetikzlibrary{arrows, arrows.meta, positioning, quotes, shapes, fit, calc, decorations.pathreplacing, calligraphy}
\tikzstyle{block} = [draw, thick, text=black, align=center, minimum width=1.2cm, minimum height=0.6cm, font=\small]
\tikzstyle{Block} = [draw, thick, rounded corners, text=black, align=center, minimum width=2cm, minimum height=1.2cm, font=\small]
\usepackage{circuitikz}
\ctikzset{bipoles/cuteswitch/thickness=0.3}
\usetikzlibrary{external,plotmarks}
\pgfplotsset{compat=newest}

\newcommand*\rectangled[1]{\tikz[baseline=(char.base)]{\node[shape=rectangle,draw,inner sep=2pt, fill=white] (char) {\footnotesize #1};}}

\begin{document}

\title{Unsupervised ANN-Based Equalizer and Its Trainable FPGA Implementation

\ifthenelse{\boolean{blind}}
{
\thanks{\centering \LARGE \textit{ Grant agreements omitted due to double-blind review}}
}
{
\thanks{This work was carried out in the framework of the CELTIC-NEXT project AI-NET-ANTILLAS (C2019/3-3) and was funded by the German Federal Ministry of Education and Research (BMBF) under grant agreements 16KIS1316 and 16KIS1317 as well as under grant 16KISK004 (Open6GHuB).}
}

}

\ifthenelse{\boolean{blind}}
{
\author{\huge \textit{Authors omitted due to double-blind review}}
}
{

\author{
\IEEEauthorblockN{Jonas Ney${}^\ddagger$,  Vincent Lauinger${}^\star$, Laurent Schmalen${}^\star$, and Norbert Wehn${}^\ddagger$}

\IEEEauthorblockA{${}^\ddagger$\textit{Microelectronic Systems Design (EMS)}, \textit{RPTU Kaiserslautern-Landau}, Germany \\
\{\texttt{ney}, \texttt{wehn}\}\texttt{@eit.uni-kl.de} \\
${}^\star$\textit{Communications Engineering Lab (CEL)}, \textit{Karlsruhe Institute of Technology (KIT)}, Germany \\
\{\texttt{vincent.lauinger}, \texttt{laurent.schmalen}\}\texttt{@kit.edu}}
}


}

\maketitle

\begin{abstract}
In recent years, communication engineers put strong emphasis on \ac{ann}-based algorithms with the aim of increasing the flexibility and autonomy of the system and its components. In this context, unsupervised training is of special interest as it enables adaptation without the overhead of transmitting pilot symbols. 
In this work, we present a novel \ac{ann}-based, unsupervised equalizer and its trainable \ac{fpga} implementation. We demonstrate that our custom loss function allows the \ac{ann} to adapt for varying channel conditions, approaching the performance of a supervised baseline. Furthermore, as a first step towards a practical communication system, we design an efficient \ac{fpga} implementation of our proposed algorithm, which achieves a throughput in the order of Gbit/s, outperforming a high-performance \acs{gpu} by a large margin.

\end{abstract}


\acresetall

\input{chapters/introduction.tex}
\input{chapters/system_model.tex}

\input{chapters/loss_function.tex}

\input{chapters/implementation.tex}
\input{chapters/results.tex}

\input{chapters/conclusion.tex}

%

\bibliography{chapters/IEEEabrv,chapters/bib_bibtex}

\end{document}

%% file: makros_own.tex
\newcommand{\ffw}{0.7\columnwidth} 
\newcommand{\ffh}{0.56\columnwidth} 

\newcommand{\Nos}{N_{\mathrm{os}}}    
\newcommand{\Nch}{N_{\mathrm{ch}}}    
\newcommand{\varN}{\sigma^2_{\mathrm{n}}} 
\newcommand{\Lf}{L_{\mathrm{fiber}}} 

\newcommand{\mlr}[1]{\mleft(#1\mright)}
\newcommand{\mlrb}[1]{\mleft\{#1\mright\}}

\DeclareMathSymbol{\mlq}{\mathord}{operators}{``}
\DeclareMathSymbol{\mrq}{\mathord}{operators}{`'}

\def\thus{\relax
	\ifmmode
	\implies
	\else
	$\implies$
	\fi}

\def\endofproof{
	\ifmmode
		\text{\hfill} \textcolor{KITgreen}{\rule{2mm}{2mm}}
	\else
		$\hfill \textcolor{KITgreen}{\rule{2mm}{2mm}}$
	\fi}

\def\j{{\mathrm{j}}}

\def\rz{\ifmmode{\mathbb{R}}%
    \else{\hbox{$\mathbb{R}$}}\fi} 
\def\nz{\ifmmode{\mathbb{N}}%
    \else{\hbox{$\mathbb{N}$}}\fi} 
\def\gz{\ifmmode{\mathbb{Z}}%
   \else{\hbox{$\mathbb{Z}$}}\fi} 
\def\cz{\ifmmode{\mathbb{C}}
    \else{\hbox{$\mathbb{C}$}}\fi}%
\def\qz{\ifmmode{\mathbb{Q}}%
    \else{\hbox{$\mathbb{Q}$}}\fi}%
\def\K{\ifmmode{\mathbb{K}}%
    \else{\hbox{$\mathbb{K}$}}\fi}%

\def\Er{\ifmmode{\mathbb{E}}%
    \else{\hbox{$\mathbb{E}$}}\fi}%
    
\def\V{
	\ifmmode
	{\mathrm{V}}
	\else
	${\mathrm{V}}$
	\fi}

\usepackage{trfsigns}

\makeatletter
\newcommand*{\AdjustMargins}{%
    \setlength{\beamer@rightmargin}{0em}%
    \setlength{\beamer@leftmargin}{0em}%
}
\makeatother

\newcommand{\hi}[2][KITyellow]{
	\ifmmode
	{
		 \mathchoice%
			{\colorbox{#1!50}{\textcolor{black}{$\displaystyle#2$}}}%
			{\colorbox{#1!50}{\textcolor{black}{$\textstyle#2$}}}
			{\colorbox{#1!50}{\textcolor{black}{$\scriptstyle#2$}}}
			{\colorbox{#1!50}{\textcolor{black}{$\scriptscriptstyle#2$}}}
	}
	\else
	{
		{\colorbox{#1!50}{\textcolor{black}{#2}}}
	}
	\fi
	}

\newcommand{\footnotetextoff}[2]{\alt<#1>{\let\thefootnote\relax\footnotetext{~}}{\footnotetext{\scriptsize #2}}}

\definecolor{kit-green100}{rgb}{0,.59,.51}
\definecolor{kit-green70}{rgb}{.3,.71,.65}
\definecolor{kit-green50}{rgb}{.50,.79,.75}
\definecolor{kit-green30}{rgb}{.69,.87,.85}
\definecolor{kit-green15}{rgb}{.85,.93,.93}
\definecolor{KITgreen}{rgb}{0,.59,.51}

\definecolor{KITpalegreen}{RGB}{130,190,60}
\colorlet{kit-maigreen100}{KITpalegreen}
\colorlet{kit-maigreen70}{KITpalegreen!70}
\colorlet{kit-maigreen50}{KITpalegreen!50}
\colorlet{kit-maigreen30}{KITpalegreen!30}
\colorlet{kit-maigreen15}{KITpalegreen!15}

\definecolor{KITblue}{rgb}{.27,.39,.66}
\definecolor{kit-blue100}{rgb}{.27,.39,.67}
\definecolor{kit-blue70}{rgb}{.49,.57,.76}
\definecolor{kit-blue50}{rgb}{.64,.69,.83}
\definecolor{kit-blue30}{rgb}{.78,.82,.9}
\definecolor{kit-blue15}{rgb}{.89,.91,.95}

\definecolor{KITyellow}{rgb}{.98,.89,0}
\definecolor{kit-yellow100}{cmyk}{0,.05,1,0}
\definecolor{kit-yellow70}{cmyk}{0,.035,.7,0}
\definecolor{kit-yellow50}{cmyk}{0,.025,.5,0}
\definecolor{kit-yellow30}{cmyk}{0,.015,.3,0}
\definecolor{kit-yellow15}{cmyk}{0,.0075,.15,0}

\definecolor{KITorange}{rgb}{.87,.60,.10}
\definecolor{kit-orange100}{cmyk}{0,.45,1,0}
\definecolor{kit-orange70}{cmyk}{0,.315,.7,0}
\definecolor{kit-orange50}{cmyk}{0,.225,.5,0}
\definecolor{kit-orange30}{cmyk}{0,.135,.3,0}
\definecolor{kit-orange15}{cmyk}{0,.0675,.15,0}

\definecolor{KITred}{rgb}{.63,.13,.13}
\definecolor{kit-red100}{cmyk}{.25,1,1,0}
\definecolor{kit-red70}{cmyk}{.175,.7,.7,0}
\definecolor{kit-red50}{cmyk}{.125,.5,.5,0}
\definecolor{kit-red30}{cmyk}{.075,.3,.3,0}
\definecolor{kit-red15}{cmyk}{.0375,.15,.15,0}

\definecolor{KITlilac}{RGB}{160,0,120}
\colorlet{kit-lila100}{KITlilac}
\colorlet{kit-lila70}{KITlilac!70}
\colorlet{kit-lila50}{KITlilac!50}
\colorlet{kit-lila30}{KITlilac!30}
\colorlet{kit-lila15}{KITlilac!15}

\definecolor{KITcyanblue}{RGB}{80,170,230}
\colorlet{kit-cyanblue100}{KITcyanblue}
\colorlet{kit-cyanblue70}{KITcyanblue!70}
\colorlet{kit-cyanblue50}{KITcyanblue!50}
\colorlet{kit-cyanblue30}{KITcyanblue!30}
\colorlet{kit-cyanblue15}{KITcyanblue!15}

\definecolor{ALUColor4}{RGB}{255,127,0}   
\definecolor{ALUColor3}{RGB}{77,175,74}   
\definecolor{ALUColor7}{RGB}{153,153,153}   
\definecolor{ALUColor2}{RGB}{55,126,184}   
\definecolor{ALUColor5}{RGB}{152,78,163}   
\definecolor{ALUColor6}{RGB}{166,86,40}    
\definecolor{ALUColor1}{RGB}{228,26,28}   

\definecolor{ALUColor1v}{RGB}{253,204,138}   
\definecolor{ALUColor2v}{RGB}{252,141,89}   
\definecolor{ALUColor3v}{RGB}{215,48,31}   

%% file: acros.tex
\acrodef{1dconv}[1D-Conv]{one-dimensional convolution}

\acrodef{ai}[AI]{artificial intelligence}
\acrodef{ask}[ASK]{amplitude-shift keying}
\acrodef{awgn}[AWGN]{additive white Gaussian noise}
\acrodef{ann}[ANN]{artificial neural network}
\acrodef{asic}[ASIC]{application-specific integrated circuit}

\acrodef{bce}[BCE]{binary cross-entropy}
\acrodef{ber}[BER]{bit error rate}
\acrodef{bler}[BLER]{block error rate}
\acrodef{bpsk}[BPSK]{binary phase-shift keying}
\acrodef{bram}[BRAM]{block random access memory}
\acrodef{bilstm}[biLSTM]{bidirectional long short-term memory}
\acrodef{bp}[BP]{backward pass}

\acrodef{cd}[CD]{chromatic dispersion}
\acrodef{cir}[CIR]{combined impulse response}
\acrodef{cma}[CMA]{constant modulus algorithm}
\acrodef{cnn}[CNN]{convolutional neural network}
\acrodef{cpu}[CPU]{central processing unit}
\acrodef{ctp}[CTP]{channel transition probability}
\acrodefplural{ctp}[CTPs]{channel transition probabilities}
\acrodef{csi}[CSI]{channel state information}

\acrodef{dnn}[DNN]{deep neural network}
\acrodef{dsp}[DSP]{digital signal processing}
\acrodef{dram}[DRAM]{dynamic random access memory}
\acrodef{dfe}[DFE]{decision-feedback equalizer}
\acrodef{dop}[DOP]{degree of paralellism}

\acrodef{fc}[FC]{fully connected}
\acrodef{fec}[FEC]{forward error correction}
\acrodef{fir}[FIR]{finite impulse response}
\acrodef{fifo}[FIFO]{first in first out}
\acrodef{fpga}[FPGA]{field programmable gate array}
\acrodef{fp}[FP]{forward pass}

\acrodef{elu}[ELU]{exponential linear unit}

\acrodef{gan}[GAN]{generative adversarial network}
\acrodef{gpu}[GPU]{graphics processing unit}

\acrodef{hls}[HLS]{high-level synthesis}
\acrodef{hwa}[HWA]{historical weight averaging}

\acrodef{imdd}[IM/DD]{intensity modulation with direct detection}
\acrodef{iot}[IoT]{Internet of things}
\acrodef{isi}[ISI]{inter-symbol interference}

\acrodef{lms}[LMS]{least-mean-square}
\acrodef{lut}[LUT]{look-up table}

\acrodef{mlsd}[MLSD]{maximum-likelihood sequence detection}
\acrodef{mse}[MSE]{mean-squared-error}
\acrodef{mac}[MAC]{multiply-accumulate}
\acrodef{ma}[MA]{moving average}
\acrodef{map}[MAP]{maximum a posteriori}

\acrodef{nn}[NN]{neural network}
\acrodef{ns}[NS]{non-saturating}

\acrodef{pam}[PAM]{pulse-amplitude modulation}
\acrodef{pdf}[pdf]{probability density function}
\acrodef{pe}[PE]{processing element}
\acrodef{pmf}[pmf]{probability mass function}

\acrodef{qlf}[QLF]{quantization loss factor}
\acrodef{rc}[RC]{raised-cosine}
\acrodef{relu}[ReLU]{rectified linear unit}
\acrodef{rls}[RLS]{recursive least squares}

\acrodef{ser}[SER]{symbol error rate}
\acrodef{sgd}[SGD]{stochastic gradient descent}
\acrodef{sld}[SLD]{square-law detection}
\acrodef{snr}[SNR]{signal-to-noise ratio}
\acrodef{sps}[sps]{samples per symbol}
\acrodef{ssmf}[SSMF]{standard single-mode fiber}
\acrodef{simd}[SIMD]{single instruction multiple data}

\acrodef{ti}[TI]{training iteration}

\acrodef{vnle}[VNLE]{Volterra-based nonlinear equalizer}

%% file: chapters/introduction.tex
\section{Introduction} \label{Sec:introduction}

The goal of next-generation communication systems is not only to increase throughput, lower latency, and improve reliability, but also to enhance autonomy by exploiting \ac{ann}-based communication algorithms \cite{wang2020}, which allow for adaptation to varying channel conditions. 
Although such algorithms often enhance the communication performance of traditional approaches \cite{zerguine2001, schaedler2019, ney2022}, the adaptation to changing conditions is based on a huge amount of data, required to perform supervised training of the \ac{ann}. This training data needs to be transmitted as pilot symbols, lowering the net throughput and information rate of the communication system.\\
To solve this problem, an \ac{ann}-based channel equalizer is proposed in \cite{lauringer2022}, which utilizes a \ac{gan} to enable unsupervised training. For unsupervised training, no labels are required, therefore it can be performed without the overhead of transmitting pilot symbols. However, the \ac{gan} approach comes with increased computational complexity and instability introduced by an additional \ac{ann} serving as loss function.

In this work, a similar approach is presented, but instead of using a discriminator \ac{ann}, the training of the \ac{ann}-based equalizer is performed using a novel low-complexity unsupervised loss function. After initial supervised training, it allows for adaptation to varying channel conditions, with the advantages of operating in a blind and channel-agnostic way.\\
However, for a practical baseband processing system, not only the communication performance but also the implementation complexity needs to be analyzed. Therefore, we present a custom hardware architecture of the unsupervised \ac{ann}-based equalizer. As hardware platform, we select \acp{fpga} as they offer arbitrary precision datatypes, custom datapaths, as well as huge bit-level parallelism. Furthermore, \acp{fpga} are highly flexible as the hardware can be reconfigured, for instance, to adapt to different application requirements.  Additionally, an \ac{fpga} design is a first step towards a custom \ac{asic} as used in practical communication systems.

In contrast to most previous works \cite{kaneda2021, li2021, freire2022},  we do not only propose an optimized implementation of the \ac{ann}'s \ac{fp} but also tackle the challenges of implementing the backpropagation algorithm on the \ac{fpga}, which enables online retraining on the edge device itself, to adapt for varying channel conditions. A related approach is also presented in \cite{Liu2023}, but contrary to our work their model is based on the split-step solution of the Manakov-PMD equation instead of an \ac{ann}, thus it is not channel-agnostic. Further, it is trained in a supervised way. 

In summary, we propose a novel unsupervised loss function and demonstrate its performance for changing channel conditions. Additionally, we present the corresponding \ac{fpga} architecture and show that it can achieve \si{\giga bit \per \second} throughput, outperforming high-end \ac{gpu} implementations.

%% file: chapters/system_model.tex
\vspace*{-2mm}
\section{System Model}
\label{Sec:system_model}

A digital communication system consists of a transmitter and a receiver with the goal of reliably transmitting information over a noisy channel. The transmitted vector $\bm{x}$, consisting of symbols $x_i\in\mathcal{A}$ from an alphabet $\mathcal{A} = \mlrb{A_1, \ldots A_M}$, is distorted by a channel and results in a received vector $\bm{y}$. At the receiver, an equalizer is applied to $\bm{y}$ which aims to revert the distortion introduced by the channel to allow for decisions $\bm{\hat{x}}$ which reliably reproduce the transmitted vector $\bm{x}$. Conventionally, the equalizer is either implemented based on a linear \ac{fir} filter or as \ac{dfe}. In our case, it is represented by a \ac{cnn}. 

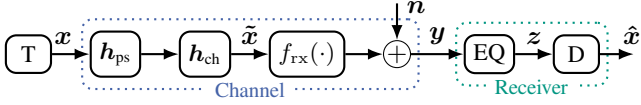
\begin{figure}[t]
	\centerline{\input{figures/tikz_system_model.tex}}
	\caption{Model of the communication chain: Data symbols $x$ are sent from a transmitter \textit{T} over a channel
        to a receiver \textit{R}.}
        \vspace*{-2mm}
	\label{fig:system_model}
\end{figure}

\subsection{Channel Model}
\label{subsec:channel_model}
As shown in Fig \ref{fig:system_model}, in our channel model, the transmitted symbols $\bm{x}$ are convolved with a \ac{rc} pulse shaping filter $\bm{h}_\mathrm{ps}$ and a linear channel impulse response $\bm{h}_\mathrm{ch}$ to produce $\bm{\tilde{x}}$. Specific receiver characteristics can be described by a possibly nonlinear function $f_\mathrm{rx}\mlr{\cdot}$. The received vector $\bm{y}$ is superimposed by a Gaussian noise vector $\bm{n}$. Finally, a decision \textit{D} is taken based on the equalizer's output $\bm{z}$. Since oversampling is essential to real systems, we run all simulations at an oversampling rate of $\Nos=2$~\ac{sps}.

We consider a dispersive optical channel with \ac{imdd} and \ac{pam} as described in \cite{plabst_wiener}. The \ac{sld} at the receiver distorts the signal nonlinearly and is modeled by
$\tilde{\bm{y}} = f_\mathrm{rx}\mlr{\tilde{\bm{x}}}$ with $\tilde{y}_i=|\tilde{x}_i|^2$. 
Linear channel distortions are caused by \ac{cd}, which can be described by its frequency response
\begin{equation*}
	H_\mathrm{cd}\left( \Lf, f \right) = \exp\left(- \frac{1}{2} \alpha \Lf \, + \, \j 2 \pi^2 \beta_2 f^2 \Lf\right),  \label{eq:cd_freqency_response}
\end{equation*}
where $\Lf$ is the fiber length, $\beta_2=-\frac{\lambda^2}{2\pi \mathrm{c}} D_\mathrm{cd}$ is defined by the wavelength $\lambda$, the speed of light $\mathrm{c}$ and the fiber's dispersion coefficient $D_\mathrm{cd}$; $\alpha$ is the fiber attenuation.
This work considers  C-band transmission at $\lambda=\SI{1550}{\nano\meter}$ over a \ac{ssmf}, with $D_\mathrm{cd} = \SI{17}{\CD}$, $\alpha \triangleq \SI{0.2}{\FL}$, and $\Lf = \SI{30}{\kilo\meter}$. Thermal noise can be modeled as \ac{awgn} with zero-mean and the variance $\varN$. For the simulation, we fix the \ac{snr} to $\SI{20}{dB}$. 
At the receiver, we carry out hard decision based on the minimum Euclidean distance.

\subsection{\ac{ann} Topology}
\label{subsec:ann_topology}

Our \ac{ann} topology is completely based on \ac{1dconv} layers, which resemble the structure of conventional digital filters. Non-linearity is introduced by \ac{relu} functions following each convolutional layer but the last. In each layer, padding is added to match the size of the output feature map to that of the input. As the data is upsampled by a factor of two before transmission, the last \ac{1dconv} layer is performed with a stride of two. Thus, one output of the \ac{cnn} corresponds to the prediction of one transmitted symbol. Our investigations have shown that a model with three \ac{1dconv} layers and a kernel size of \num{21} is sufficient. A larger model did not lead to significant gains in communication performance.




%% file: figures/tikz_system_model.tex
\begin{tikzpicture}[node distance=0.2,>=latex, scale=0.9]
    \pgfdeclarelayer{background}
    \pgfdeclarelayer{foreground}
    \pgfsetlayers{main,foreground}

    \def\minWid{0.6cm}; \def\minHei{0.5cm}; \def\arrowLen{0.5cm};
    \def\boxFontSize{\footnotesize}
    
    \def\colLight{15};
    \def\boxCol{KITgreen}; \def\DotBoxCol{KITblue};

    \begin{pgfonlayer}{foreground}
    \node[overlay] (in) {};
    \node[block, rounded corners, minimum width=\minWid, minimum height=\minHei] (transmitter) {T};

	\node[block, rounded corners, right=\arrowLen of transmitter, minimum width=\minWid] (txfilter) {$\bm{h}_\text{ps}$};
	\node[block, rounded corners, right=\arrowLen of txfilter, minimum width=\minWid] (channel) {$\bm{h}_\text{ch}$};
	\node[block, rounded corners, right=\arrowLen of channel, minimum width=\minWid] (sld) {$f_{\mathrm{rx}}\mlr{\cdot}$};
	\node[draw, circle,inner sep=-0.0pt, right=\arrowLen of sld] (noise) {$\mathbf{+}$};

    \node[block, rounded corners, minimum width=\minWid, minimum height=\minHei, right=\arrowLen+0.2cm of noise] (equalizer) {EQ};
    \node[block, rounded corners, minimum width=\minWid, minimum height=\minHei, right=\arrowLen of equalizer] (decision) {D};

	\draw[-{Latex[length=2mm]}, thick] (transmitter) -- node[midway, above, xshift=-0.1cm] {$\bm{x}$} (txfilter);
	\draw[-{Latex[length=2mm]}, thick] (txfilter) -- (channel);
	\draw[-{Latex[length=2mm]}, thick] (channel) -- node[midway, above] {$\bm{\tilde{x}}$} (sld);
	\draw[-{Latex[length=2mm]}, thick] (sld) -- (noise);
	\draw[{Latex[length=2mm]}-, thick] (noise) -- node[midway, right, yshift=+0.2cm] {$\bm{n}$} +(0,1.5*\arrowLen);
	\draw[-{Latex[length=2mm]}, thick] (noise.east) -- node[midway, above] {$\bm{y}$}  (equalizer.west);
	\draw[-{Latex[length=2mm]}, thick] (equalizer) -- node[midway, above] {$\bm{z}$} (decision);
    \draw[-{Latex[length=2mm]}, thick] (decision.east) -- node[midway, above, xshift=+0.2cm] {$\bm{\hat{x}}$} +(\arrowLen,0);
    \end{pgfonlayer}

    \node[draw, KITblue, thick, dotted, rounded corners, inner xsep=0.1cm, inner ysep=0.15cm, fit=(txfilter) (channel) (sld) (noise)] (total_channel) {};
    \node[fill=white] at (total_channel.south) {\textcolor{KITblue}{\footnotesize Channel}};

    \node[draw, KITgreen, thick, dotted, rounded corners, inner xsep=0.1cm, inner ysep=0.15cm, fit=(equalizer) (decision)] (receiver) {};
    \node[fill=white] at (receiver.south) {\textcolor{KITgreen}{\footnotesize Receiver}};
    
\end{tikzpicture}

%% file: chapters/loss_function.tex
\section{Loss Function}
\label{sec:loss_function}

For training the \ac{ann}, we use a two-step approach. First, initial training for a channel model is performed in a PyTorch environment based on supervised \ac{mse} loss. In a second step, we perform retraining of the \ac{ann} on the edge device itself, to adapt to varying channel conditions. This retraining can either be performed with supervised \ac{mse} loss or using a custom unsupervised loss function.

The main purpose of the novel, unsupervised heuristic loss function is to enable adaptation of the \ac{cnn} without any pilot symbols, resulting in less overhead and a higher net data rate. Since the unsupervised loss function does not consider the mapping of the output to the correct symbol, but only operates on the statics of the channel output, the supervised loss function is used for initial training. 

\subsection{Novel Unsupervised Loss Function}

As a first step, we show how our unsupervised loss function can be applied to \ac{pam}-2 modulation. Subsequently, we propose a way to adapt it to \ac{pam}-4 modulation. 

\subsubsection{\ac{pam}-2 modulation}
\begin{figure}[!h]
	\centerline{\input{figures/tikz_custom_loss.tex}}
	\caption{Polynomial function, with minima at constellation points $A_1$ and $A_2$.}
	\label{fig:custom_loss}
\end{figure}
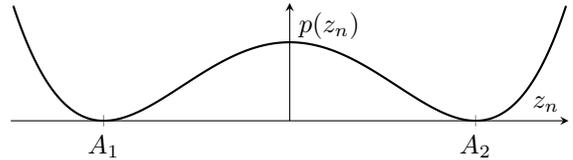

The unsupervised loss function is comprised of two parts, $\mathrm{loss}_a(\bm{z})$ and $\mathrm{loss}_b(\bm{z})$. First, a polynomial function $p(\cdot)$ is used to push each of the \ac{cnn}'s outputs to the actual constellation points $A_1$ and $A_2$:
\begin{equation*}
    p(z_n) = (z_n-A_1)^2 \cdot (z_n-A_2)^2 \; .
\end{equation*}
Then $\mathrm{loss}_a(\bm{z})$ is given as the sum of $p(z_n)$ over all outputs $z_n$ for a sequence of length $N$:
\vspace*{-2mm}
\begin{equation*}
    \mathrm{loss}_a(\bm{z})= \sum_{n=1}^{N} p(z_n) \; .
\end{equation*}
As shown in Fig. \ref{fig:custom_loss}, $p(z_n)$ has global minima at $A_1$ and $A_2$ which correspond to the possible input symbols of the transmitted vector $\bm{x}$. Thus, the outputs of the \ac{cnn} are pushed to one of the undistorted symbols. Since the \ac{cnn} is initially trained in a supervised fashion, the channel output $z_i$ is mapped to the actual transmitted symbol $x_i$ at the beginning of the unsupervised training. We expected that $\mathrm{loss}_a(\bm{z})$ forces the network to keep $z_i$ at the corresponding transmitted symbol $x_i$ even if the channel changes during unsupervised training. 

However, we observed that during training all received symbols $\bm{z}$ were either pushed to the minimum $A_1$ or to the minimum $A_2$. Thus we propose a second loss function $\mathrm{loss}_b(\bm{z})$ which forces $\bm{z}$ to be equally spread between $A_1$ and $A_2$. Therefore, we define $d_i$ as the accumulated absolute distance of each output of one sequence with length $N$ to the constellation point $A_i$:
\vspace*{-2mm}
\begin{equation*}
    d_i=\sum_{n=1}^{N} |z_n - A_i| \; .
\end{equation*}
Subsequently, we determine the absolute difference of the distances to $A_1$ and $A_2$, as
\begin{equation*}
    \mathrm{loss}_b(\bm{z})=|d_1 - d_2| \; .
\end{equation*}
Thus $\mathrm{loss}_b(\bm{z})$ is minimal for $d_1 = d_2$, which is the case when all outputs $z_n$ are equally distributed between $A_1$ and $A_2$.
Finally, $\mathrm{loss}(\bm{z})$ is calculated as
\begin{equation*}
    \mathrm{loss}(\bm{z})= \mathrm{loss}_a(\bm{z}) + \mu \cdot \mathrm{loss}_b(\bm{z})\; ,
\end{equation*}
where $\mu$ is a weighting factor to balance between $\mathrm{loss}_a(\bm{z})$ and $\mathrm{loss}_b(\bm{z})$ ($\mu$ is set to \num{4} in our experiments).

To summarize, our heuristic unsupervised loss function forces the equalizer output to be close to the constellation symbols and equally spread between them. Since the constellation symbols represent the channel input, the equalizer learns to resemble this input at its output. Therefore, by unsupervised training, the outputs are prevented from drifting during changing channel conditions, resulting in an increased communication performance, as shown in Sec. \ref{Sec:results}


\subsubsection{PAM-4 modulation}

In addition to the \ac{pam}-2 based loss function, we demonstrate how the unsupervised loss function can be adapted to higher-order modulation schemes e.g. \ac{pam}-4.
The first part of the loss function $\mathrm{loss}_a(\bm{z})$ is similar to the \ac{pam}-2 example, but with minima at each of the four constellation points $A_1, A_2, A_3, A_4$: 
\begin{equation*}
    \mathrm{loss}_a(\bm{z})=\sum_{n=1}^{N}\prod_{i=1}^{4}(z_n-A_i)^2 \; .
\end{equation*}

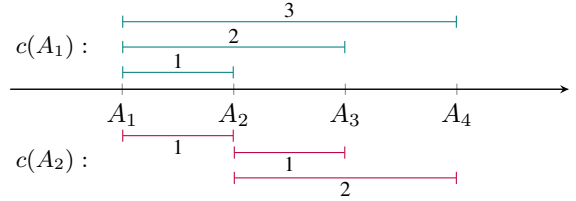
\begin{figure}[t]
	\centerline{\input{figures/tikz_4pam_illustration.tex}}
        \vspace*{-2mm}
	\caption{Illustration of $c(A_1)$ and $c(A_2)$ for \ac{pam}-4, with a distance of one between each constellation point.} 
        \vspace*{-5mm}	
        \label{fig:pam4_c_illustration}
\end{figure}

In contrast, a modification is needed for $\mathrm{loss}_b(\bm{z})$. As explained previously, $\mathrm{loss}_b(\bm{z})$ is introduced to equally distribute all outputs $\bm{z}$ across the constellation points based on the accumulated distance of each output to each constellation point. 
For \ac{pam}-4, $\mathrm{loss}_b(\bm{z})$ is constructed based on four distances $d_i$, one for each constellation point $A_i$. Further, we define $c(A_i)$ as the distance of $A_i$ to the remaining constellation points: 
\vspace*{-2mm}
\begin{equation*}
    c(A_i) = \sum_{\substack{j=1 \ j \neq i}}^{4}(|A_i - A_j|)
\end{equation*}
Then $c(A_2) = c(A_3) = 4$ and $c(A_1) = c(A_4) = 6$, as shown in Fig. \ref{fig:pam4_c_illustration}. Thus, for compensation $d_2$ and $d_3$ are multiplied by $\frac{3}{2}$:
\vspace*{-2mm}
\begin{equation*}
    \mathrm{loss}_b(\bm{z}) = \Bigl| d_1 - d_4\Bigr| + \Bigl|\frac{3}{2} d_2 - \frac{3}{2} d_3\Bigr| + \Bigl|d_1 - \frac{3}{2} d_2\Bigr| + \Bigl|d_4 - \frac{3}{2} d_3\Bigr| \; . 
\end{equation*}
This way, $\mathrm{loss}_b(\bm{z})$ is minimal if the outputs of the network are equally distributed between $A_1$ to $A_4$.
Similar to \ac{pam}-2, the final loss is given as the sum of $\mathrm{loss}_a(\bm{z})$ and $\mathrm{loss}_b(\bm{z})$ with a weighting factor $\mu$.

%% file: figures/tikz_custom_loss.tex
\begin{tikzpicture}[node distance=0.3,>=latex]
    \tikzset{near start abs/.style={xshift=.01cm}}

    \begin{axis}[
        	restrict y to domain=0:1.5, 
            xmin=-1.5, 
            xmax=1.5, 
            ymin=0, 
            ymax=1.5, 
            axis y line=center,
            axis x line=middle,
            samples=1000,
            xtick={-1, 1},
            xticklabels={$A_1$, $A_2$},
            ytick=\empty,
            xlabel={$z_n$},
            ylabel={$p(z_n)$},
            width=\columnwidth,
            height=\columnwidth*0.35
            ]
      \addplot[black, thick] (x, x*x*x*x-2*x*x+1);
    \end{axis}
    
\end{tikzpicture}

%% file: figures/tikz_4pam_illustration.tex
\begin{tikzpicture}[node distance=0.3,>=latex]
    
  \begin{axis}[
      axis x line=middle,
      axis y line=none,
      xmin=0,
      xmax=5,
      xticklabels={$A_1$, $A_2$, $A_3$, $A_4$},
      xtick={1,2,3,4},
      width=\columnwidth,
      height=\columnwidth*0.8
    ]

\node at (axis cs:0.37,0.2) {\small$c(A_1):$};
\draw[|-|, teal] (axis cs:1,0.08) -- node [yshift=-0.7mm, above] {\footnotesize \textcolor{black}{1}} (axis cs:2,0.08);   
\draw[|-|, teal] (axis cs:1,0.2) -- node [yshift=-0.7mm, above] {\footnotesize \textcolor{black}{2}} (axis cs:3,0.2);   
\draw[|-|, teal] (axis cs:1,0.32) -- node [yshift=-0.7mm, above] {\footnotesize \textcolor{black}{3}} (axis cs:4,0.32);

\node at (axis cs:0.37,-0.34) {\small$c(A_2):$};
\draw[|-|, purple] (axis cs:1,-0.22) -- node [yshift=0.7mm, below] {\footnotesize \textcolor{black}{1}} (axis cs:2,-0.22);   
\draw[|-|, purple] (axis cs:2,-0.3) -- node [yshift=0.7mm, below] {\footnotesize \textcolor{black}{1}} (axis cs:3,-0.3);   
\draw[|-|, purple] (axis cs:2,-0.42) -- node [yshift=0.7mm, below] {\footnotesize \textcolor{black}{2}} (axis cs:4,-0.42);

\addplot[line width=0.1pt, mark=none]{0};

    \end{axis}
    
\end{tikzpicture}

%% file: chapters/implementation.tex
\section{Implementation}

As a first step towards a practical communication system, we present an efficient \ac{fpga} implementation of our proposed algorithm. 

\subsection{Quantization}
\label{subsec:quantization}




For minimizing the implementation complexity, all values of the \ac{cnn}'s \ac{fp} and \ac{bp} are represented as quantized fixed-point numbers. To find the optimal bit width for each layer, we perform an in-depth quantization analysis for weights, activations, multipliers, accumulators, and gradients.


In a first step, we select an appropriate quantization scheme for weights and activations by adapting the automatic quantization strategy proposed in \cite{nikolic2020}. Therefore, the loss function is modified to simultaneously learn the precision of each layer while optimizing the accuracy of the \ac{ann} during training. This is achieved by using a differentiable interpolation of the bit-widths, which allows to train them using backpropagation. Similar to \cite{nikolic2020}, we include a trade-off factor in the loss function, which determines how aggressively to quantize. This enables efficient exploration of the trade-off between bit width and communication performance. 


\begin{figure}[b]
	\centering
	\input{figures/tikz_quant_sweep_results.tex}
	\caption{Plot of the average number of bits used for weights and activations vs \ac{ber} achieved with this quantization. The dotted line represents the Pareto front, connecting all points with the best trade-off between complexity and accuracy. The red square marks the final model used for implementation.}
	\label{fig:quant_sweep}
\end{figure}
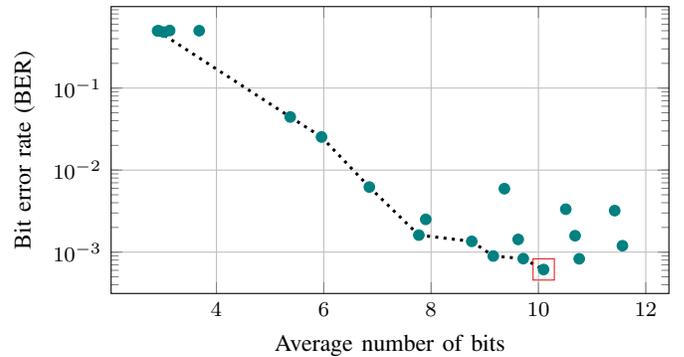

Fig. \ref{fig:quant_sweep} shows the quantization analysis for the \ac{pam}-2 channel with a symbol rate of \SI{25}{\giga Bd} and the model described in \ref{subsec:ann_topology}. The dotted line connects the Pareto optimal points, where each is trained with a different trade-off factor. For implementation, we selected the model marked by the red square, as it achieves a low \acs{ber} (\num{0.0006}) with moderate complexity (\num{10.1} bits in average). Moreover, a higher number of bits didn't show any improvement in communication performance.

In contrast to the quantization of weights and activations, the quantization of accumulators, multipliers, and gradients can't be learned during training. Therefore, we simulate inference and training for multiple randomly generated input sequences. We select the integer and decimal bits for each datatype, as the minimal number of bits needed to cover the whole dynamic range of values profiled in our simulation.

\subsection{Hardware Architecture}
\label{subsec:hardware_architecture}

\begin{figure}[b]
	\centerline{\input{figures/tikz_hardware_architecture.tex}}
	\caption{Hardware architecture with kernel size $K$, padding $P$, stride $S$, dilation $D$ and number of channels $N_{ch}$. Arrows indicate streams of feature maps, either between subsequent layers or between \ac{fp} and \ac{bp}. Feature map buffers are shown in red.}
  \label{fig:hardware_architecture}
\end{figure}
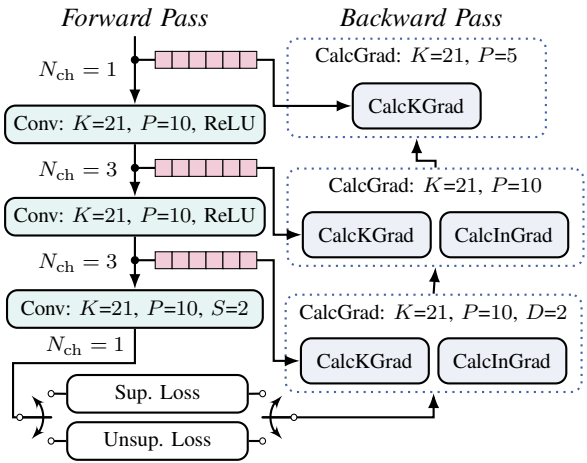

A major constraint for high-performance hardware implementation of \acp{ann} is the restricted on-chip memory of the \ac{fpga}. The quantization presented in Sec.~\ref{subsec:quantization} is a first step to reduce the memory footprint. However, in contrast to most previous works, we also implement the training of the \ac{ann} on the \ac{fpga}, which requires additional resources. To perform backpropagation, a large amount of memory is required to store the feature maps of the \ac{fp} to be reused in the \ac{bp}. For large sequence lengths, the \ac{fpga}'s on-chip memory is usually not sufficient to store those feature maps, thus offloading to external DRAM is necessary, resulting in a limited throughput. In particular, for an implementation with a sequence length of five Ethernet packages, the feature map buffers of one \ac{cnn} instance nearly consume \SI{50}{\percent} of the \ac{bram} resources of the \textit{Xilinx ZCU102}. This limits the memory available for the network's weights as well as the achievable throughput, as fewer \ac{cnn} instances can be placed on the board.\\
To solve this problem, we propose a fully pipelined architecture in which we balance the lifetime of the feature maps such that the memory footprint is reduced, as sketched in Fig.~\ref{fig:hardware_architecture}. 

\begingroup
\abovedisplayskip=8pt 
\belowdisplayskip=8pt

In our hardware architecture, all modules of the \ac{fp} (shown in green) and the \ac{bp} (shown in blue) are implemented as separate pipeline stages. The forward blocks (Conv) calculate the discrete convolution of input $\bm{i}$ with kernel $\bm{k}$ and apply the \ac{relu} activation function
\begin{equation}
\label{eq_1}
\bm{o} = \mathrm{ReLU}(\bm{i} \ast \bm{k}) \; .
\end{equation}
The backward blocks (CalcGrad) consist of two modules: one to calculate the input's gradient (CalcInGrad) and another one to calculate the kernel's gradient (CalcKGrad). The input's gradient $\nabla \bm{i}$ is given as the channel-wise convolution of the flipped kernel with the output gradient $\nabla \bm{o}$:
\begin{equation}
\label{eq_2}
\nabla \bm{i} = \mathrm{flip}(\bm{k}) \ast \nabla \bm{o} \; ,
\end{equation}
while the kernel gradient $\nabla \bm{k}$ can be obtained by convolving the input with the output gradient:
\begin{equation}
\label{eq_3}
\nabla \bm{k} = \bm{i} \ast \nabla \bm{o} \; .
\end{equation}
\endgroup
The core of our hardware architecture is a highly customizable convolutional module, which allows for variable kernel size, padding, stride, and dilation. Each convolution of the \ac{cnn} is implemented as a separate hardware module, enabling parallel computation of each layer and therefore increasing the overall throughput, which is crucial for high-performance communication systems.\\
To solve the problem of large feature map buffers between the \ac{fp} and \ac{bp}, we take advantage of the sequential nature of convolution operations \eqref{eq_2} and \eqref{eq_3} in the \ac{bp}. The calculation of $\nabla \bm{i}$ and $\nabla \bm{k}$ can already start before all elements of $\bm{i}$ and $\nabla \bm{o}$ are available. Since the \ac{fp} and the \ac{bp} are implemented as separate pipeline stages, they can be processed in parallel. Therefore, the feature maps are read and written concurrently and the size of the buffers can be greatly reduced.
Moreover, we designed our layers in such a way that \ac{fp} and \ac{bp} have similar latency. Hence, the size of the feature map buffers is determined only by the depth of the pipeline and does not depend on the length of the sequence. This enables processing symbol sequences of arbitrary length. Thus our approach is suited for many different application scenarios.\\
To increase flexibility even further, our architecture enables variable \ac{dop} on the level of input channels, output channels, kernel size, and the number of instances. This way, the number of \ac{mac} operations performed per clock cycle can be adjusted as required. On the one hand, this enables optimizing the utilization of available hardware resources and thus increases efficiency. On the other hand, variable \ac{dop} allows to trade-off power consumption against throughput to adapt to specific application requirements by reconfiguration of the \ac{fpga}.

%% file: figures/tikz_quant_sweep_results.tex
\begin{tikzpicture}[
        square/.style={regular polygon,regular polygon sides=4},
        tight background,
	/pgfplots/every mark/.append style={solid, mark size=3.5pt},
	/pgfplots/every axis/.append style={font=\small}, 
	/pgfplots/every tick label/.append style={font=\footnotesize},
	/pgfplots/every axis legend/.append style={nodes={right},font={\footnotesize},row sep=-2pt}  
	]
	\begin{axis}[
            ymode=log, 
            xlabel={Average number of bits},
            ylabel={Bit error rate (BER)},
            width=\columnwidth,
            height=\columnwidth*0.6,
            xmajorgrids,
	    ymajorgrids
            ]


    \draw[dotted, line width=0.4mm] (axis cs:3.0182453,0.48106334) -- (axis cs:2.903856119,0.49763044);
    \draw[dotted, line width=0.4mm] (axis cs:2.903856119,0.49763044) -- (axis cs:5.375129382,0.044404621);
    \draw[dotted, line width=0.4mm] (axis cs:5.375129382,0.044404621) -- (axis cs:5.95807449,0.025328555);
    \draw[dotted, line width=0.4mm] (axis cs:5.95807449,0.025328555) -- (axis cs:6.849249363,0.006212664);
    \draw[dotted, line width=0.4mm] (axis cs:6.849249363,0.006212664) -- (axis cs:7.771092097,0.001611244);
    \draw[dotted, line width=0.4mm] (axis cs:7.771092097,0.001611244) -- (axis cs:8.75763464,0.001352715);
    \draw[dotted, line width=0.4mm] (axis cs:8.75763464,0.001352715) -- (axis cs:9.158255736,0.000896626);
    \draw[dotted, line width=0.4mm] (axis cs:9.158255736,0.000896626) -- (axis cs:9.717003028,0.000833001);
    \draw[dotted, line width=0.4mm] (axis cs:9.717003028,0.000833001)-- (axis cs:10.0960145,0.000612859);

    \addplot [only marks, mark=*, mark options={teal, scale=1}] table [x=Avg_Bits,y=BER]{data/Quantization/pam2_long.txt}; 
    \coordinate (Center) at (axis cs:10.096,0.000613);
    
    \end{axis}

    \node at (Center) [square, draw, ALUColor1, inner sep=1mm] (square_node) {};

\end{tikzpicture}

%% file: figures/tikz_hardware_architecture.tex
\begin{tikzpicture}[node distance=0.2,>=latex, scale=0.9]
    \tikzset{near start abs/.style={xshift=.01cm}};
    \def\minWid{3.4cm}; \def\minHei{0.5cm}; \def\arrowLen{0.7cm};
    \def\boxFontSize{\footnotesize}
    
    \def\colLight{10};
    \def\boxCol{KITgreen}; \def\DotBoxCol{KITblue};


    \node[text=black] (forward_text) {\textit{Forward Pass}};
    \node[right=1.5cm of forward_text, text=black] (backward_text) {\textit{Backward Pass}};


    \node[block, draw=black, rounded corners, fill=KITgreen!\colLight, minimum width=\minWid, minimum height=\minHei, below=\arrowLen*1.3 of forward_text] (conv0) {\boxFontSize Conv: $K$=$21$, $P$=$10$, ReLU};

    \node[block, draw=black, rounded corners, fill=KITgreen!\colLight, minimum width=\minWid, minimum height=\minHei, below=\arrowLen of conv0] (conv1) {\boxFontSize Conv: $K$=$21$, $P$=$10$, ReLU};

    \node[block, draw=black, rounded corners, fill=KITgreen!\colLight, minimum width=\minWid, minimum height=\minHei, below=\arrowLen of conv1] (conv2) {\boxFontSize Conv: $K$=$21$, $P$=$10$, $S$=$2$};

    \node[block, draw=black, rounded corners, minimum width=\minWid - 1cm, minimum height=\minHei, below=0.6cm of conv2, xshift=0.3cm] (sup_loss) {\boxFontSize Sup. Loss};
    \node[block, draw=black, rounded corners, minimum width=\minWid - 1cm, minimum height=\minHei, below=0.1cm of sup_loss] (unsup_loss) {\boxFontSize Unsup. Loss};

    \node[draw, above right=-0.264cm and 0.6cm of unsup_loss, cute spdt mid arrow, xscale=-0.7, yscale=0.75] (loss_switch_right) {};
    \node[draw, above left=-0.264cm and 0.2cm of unsup_loss, cute spdt mid arrow, xscale=0.7, yscale=0.75] (loss_switch_left) {};

    \node[right=0.3cm of conv2, minimum width=\minWid+0.5cm, yshift=-0.05cm] (grad2) {\boxFontSize CalcGrad: $K$=$21$, $P$=$10$, $D$=$2$};
    \node[block, draw=black, rounded corners, fill=KITblue!\colLight, minimum width=\minWid*0.5, left=0cm of grad2.south, xshift=-0.05cm, yshift=-0.4cm] (k_grad2) {\boxFontSize CalcKGrad};
    \node[block, draw=black, rounded corners, fill=KITblue!\colLight, minimum width=\minWid*0.5, right=0cm of grad2.south, xshift=0.05cm, yshift=-0.4cm] (in_grad2) {\boxFontSize CalcInGrad};
    \node[draw, \DotBoxCol, thick, dotted, rounded corners, inner xsep=0.01cm, inner ysep=0.05cm, yshift=-0.05cm, fit=(grad2) (k_grad2) (in_grad2)] (box_bp_2) {};

    \node[right=0.3cm of conv1, minimum width=\minWid+0.5cm, yshift=0.4cm] (grad1) {\boxFontSize CalcGrad: $K$=$21$, $P$=$10$};
    \node[block, draw=black, rounded corners, fill=KITblue!\colLight, minimum width=\minWid*0.5, left=0cm of grad1.south, xshift=-0.05cm, yshift=-0.4cm] (k_grad1) {\boxFontSize CalcKGrad};
    \node[block, draw=black, rounded corners, fill=KITblue!\colLight, minimum width=\minWid*0.5, right=0cm of grad1.south, xshift=0.05cm, yshift=-0.4cm] (in_grad1) {\boxFontSize CalcInGrad};
    \node[draw, \DotBoxCol, thick, dotted, rounded corners, inner xsep=0.01cm, inner ysep=0.05cm, yshift=-0.05cm, fit=(grad1) (k_grad1) (in_grad1)] (box_bp_1) {};

    \node[right=0.3cm of conv0, minimum width=\minWid, yshift=0.9cm] (grad0) {\boxFontSize CalcGrad: $K$=$21$, $P$=$5$};
    \node[block, draw=black, rounded corners, fill=KITblue!\colLight, minimum width=\minWid*0.5, below=0.1cm of grad0] (k_grad0) {\boxFontSize CalcKGrad};
    \node[draw, \DotBoxCol, thick, dotted, rounded corners, inner xsep=0.01cm, inner ysep=0.05cm, yshift=-0.05cm, fit=(grad0) (k_grad0)] (box_bp_0) {};

    \draw[-{Latex[length=2mm]}, thick] (forward_text) -- node[midway, left, xshift=-0.1cm] {\footnotesize $\Nch=1$} (conv0.north);
    \filldraw (forward_text.south) +(0,-\arrowLen*0.5) circle (1.5pt);

    \draw[-{Latex[length=2mm]}, thick] (conv0) -- node[midway, left, xshift=-0.1cm] {\footnotesize $\Nch=3$} (conv1.north);
    \filldraw (conv0.south) +(0,-\arrowLen*0.5) circle (1.5pt);

    \draw[-{Latex[length=2mm]}, thick] (conv1) -- node[midway, left, xshift=-0.1cm] {\footnotesize $\Nch=3$} (conv2.north);
    \filldraw (conv1.south) +(0,-\arrowLen*0.5) circle (1.5pt);

    \draw[-{Latex[length=2mm]}, thick] (forward_text.south)++(0, -\arrowLen*0.5) -- +(2.05cm, 0) |- (k_grad0.west);
    \draw[-{Latex[length=2mm]}, thick] (conv0.south)++(0, -\arrowLen*0.5) -- +(2.05cm, 0) |- (k_grad1.west);
    \draw[-{Latex[length=2mm]}, thick] (conv1.south)++(0, -\arrowLen*0.5) -- +(2.05cm, 0) |- (k_grad2.west);

    \def\recSize{0.25cm};
    
    \filldraw[draw=black, fill=purple!20] (forward_text.south)++(0.3cm, -\arrowLen*0.5 + \recSize * 0.5) rectangle ++(\recSize,-\recSize);
    \filldraw[draw=black, fill=purple!20] (forward_text.south)++(0.3cm + 1 * \recSize, -\arrowLen*0.5 + \recSize * 0.5) rectangle ++(\recSize,-\recSize);
    \filldraw[draw=black, fill=purple!20] (forward_text.south)++(0.3cm + 2 * \recSize, -\arrowLen*0.5 + \recSize * 0.5) rectangle ++(\recSize,-\recSize);
    \filldraw[draw=black, fill=purple!20] (forward_text.south)++(0.3cm + 3 * \recSize, -\arrowLen*0.5 + \recSize * 0.5) rectangle ++(\recSize,-\recSize);
    \filldraw[draw=black, fill=purple!20] (forward_text.south)++(0.3cm + 4 * \recSize, -\arrowLen*0.5 + \recSize * 0.5) rectangle ++(\recSize,-\recSize);
    \filldraw[draw=black, fill=purple!20] (forward_text.south)++(0.3cm + 5 * \recSize, -\arrowLen*0.5 + \recSize * 0.5) rectangle ++(\recSize,-\recSize);

    \filldraw[draw=black, fill=purple!20] (conv0.south)++(0.3cm, -\arrowLen*0.5 + \recSize * 0.5) rectangle ++(\recSize,-\recSize);
    \filldraw[draw=black, fill=purple!20] (conv0.south)++(0.3cm + 1 * \recSize, -\arrowLen*0.5 + \recSize * 0.5) rectangle ++(\recSize,-\recSize);
    \filldraw[draw=black, fill=purple!20] (conv0.south)++(0.3cm + 2 * \recSize, -\arrowLen*0.5 + \recSize * 0.5) rectangle ++(\recSize,-\recSize);
    \filldraw[draw=black, fill=purple!20] (conv0.south)++(0.3cm + 3 * \recSize, -\arrowLen*0.5 + \recSize * 0.5) rectangle ++(\recSize,-\recSize);
    \filldraw[draw=black, fill=purple!20] (conv0.south)++(0.3cm + 4 * \recSize, -\arrowLen*0.5 + \recSize * 0.5) rectangle ++(\recSize,-\recSize);
    \filldraw[draw=black, fill=purple!20] (conv0.south)++(0.3cm + 5 * \recSize, -\arrowLen*0.5 + \recSize * 0.5) rectangle ++(\recSize,-\recSize);

    \filldraw[draw=black, fill=purple!20] (conv1.south)++(0.3cm, -\arrowLen*0.5 + \recSize * 0.5) rectangle ++(\recSize,-\recSize);
    \filldraw[draw=black, fill=purple!20] (conv1.south)++(0.3cm + 1 * \recSize, -\arrowLen*0.5 + \recSize * 0.5) rectangle ++(\recSize,-\recSize);
    \filldraw[draw=black, fill=purple!20] (conv1.south)++(0.3cm + 2 * \recSize, -\arrowLen*0.5 + \recSize * 0.5) rectangle ++(\recSize,-\recSize);
    \filldraw[draw=black, fill=purple!20] (conv1.south)++(0.3cm + 3 * \recSize, -\arrowLen*0.5 + \recSize * 0.5) rectangle ++(\recSize,-\recSize);
    \filldraw[draw=black, fill=purple!20] (conv1.south)++(0.3cm + 4 * \recSize, -\arrowLen*0.5 + \recSize * 0.5) rectangle ++(\recSize,-\recSize);
    \filldraw[draw=black, fill=purple!20] (conv1.south)++(0.3cm + 5 * \recSize, -\arrowLen*0.5 + \recSize * 0.5) rectangle ++(\recSize,-\recSize);

    \draw[-, thick] (conv2) |- node[near start, left] {\footnotesize $\Nch=1$} +(-1.8cm, -0.8cm) |- (loss_switch_left.in);
    \draw[-, thick] (loss_switch_left.out 1) -- (sup_loss.west);
    \draw[-, thick] (loss_switch_left.out 2) -- (unsup_loss.west);

    \draw[-{Latex[length=2mm]}, thick] (loss_switch_right.in) -| (box_bp_2.south);
    \draw[-, thick] (loss_switch_right.out 1) -- (sup_loss.east);
    \draw[-, thick] (loss_switch_right.out 2) -- (unsup_loss.east);


    \draw[-{Latex[length=2mm]}, thick] (box_bp_2.north) -- (box_bp_1.south);

    \draw[-{Latex[length=2mm]}, thick] (box_bp_1.north) -- +(0,0.1cm) -| (box_bp_0.south);

\end{tikzpicture}

%% file: chapters/results.tex
\section{Results}
\label{Sec:results}
The following results are evaluated based on the channel described in Sec.~\ref{subsec:channel_model} with the \ac{ann} topology of Sec.~\ref{subsec:ann_topology}. 



\subsection{Adaptation Analysis}

\begin{figure}[h]
	\centering
	\input{figures/tikz_retraining_dispersion.tex}
        \vspace*{-4mm}
	\caption{Results for the \ac{pam}-2 channel with a symbol rate of \SI{25}{\giga Bd}. Initial training is performed for $D_\mathrm{cd}=\SI{17}{\CD}$. $\mathrm{CNN}_{\mathrm{sup.}}$ and $\mathrm{CNN}_{\mathrm{unsup.}}$ are retrained in steps of \SI{1.8}{\CD}.}
	\label{fig:retraining_dispersion}
\end{figure}

\begin{figure}[h]
	\centering
	\input{figures/tikz_retraining_dispersion_pam4.tex}
        \vspace*{-4mm}
	\caption{Results for the \ac{pam}-4 channel with a symbol rate of \SI{20}{\giga Bd} for different $D_\mathrm{cd}$. Initial training is performed for $D_\mathrm{cd}=\SI{17}{\CD}$. $\mathrm{CNN}_{\mathrm{sup.}}$ and $\mathrm{CNN}_{\mathrm{unsup.}}$ are retrained in steps of \SI{1.8}{\CD}.}
	\label{fig:retraining_dispersion_pam4}
\end{figure}

As the main purpose of our approach is the adaptation to varying channel conditions on the edge device, we evaluate how supervised and unsupervised retraining of the \ac{cnn}-based equalizer improves the communication performance. The baseline of our analysis corresponds to a model that is trained from scratch for every new channel condition. Further, we give results for a model which is only trained for the initial channel but not retrained for the changing conditions. We also show the \ac{ber} for unsupervised and supervised retraining of the \ac{cnn}, performed during the channel variation. Those models are retrained for \num{500} iterations with a learning rate of \num{0.02} with \ac{sgd}. 
We also evaluate the \ac{ber} of a third-order Volterra equalizer~\cite{stojanovic2017volterra} with memory $F=[35, 17, 9]$ and, for fair comparison, approximately the same number of parameters as our \ac{cnn}, which is trained in a supervised way based on the \ac{mse} loss. As a varying channel characteristic, we select the fiber dispersion parameter $D_\mathrm{cd}$. This property of the optical fiber may change due to temperature, aging effects, and other environmental factors.\\
The results are shown in Fig. \ref{fig:retraining_dispersion} and Fig. \ref{fig:retraining_dispersion_pam4} for \ac{pam}\nobreakdash-2 and \ac{pam}\nobreakdash-4 respectively. As expected, for both cases the \ac{ber} increases significantly for high $D_\mathrm{cd}$ if no retraining is performed. Especially the gap to the baseline, which is trained from scratch, grows dramatically. However, by retraining the model in unsupervised fashion, the gap to the baseline can be highly reduced. Specifically, it is decreased by a factor of \num{6} for $D_\mathrm{cd}=\SI{26}{\CD}$ for \ac{pam}\nobreakdash-2, and by a factor of \num{2} for \ac{pam}\nobreakdash-4. For \ac{pam}\nobreakdash-2 the performance of unsupervised retraining is similar to the supervised one, whereas supervised retraining has a slightly better performance for \ac{pam}\nobreakdash-4. This indicates that our unsupervised loss function is well suited for performing adaptation to varying channel conditions, especially for \ac{pam}\nobreakdash-2. In contrast to the supervised loss function, no labeled training data in form of pilot symbols is required, which increases the overall information rate.\\
Moreover, both retraining techniques outperform the conventional, supervised Volterra equalizer over the whole range of $D_\mathrm{cd}$ for \ac{pam}\nobreakdash-2 and \ac{pam}\nobreakdash-4, validating the potential of \ac{ann}-based equalization from a communication perspective. 

\begin{figure}[t]
	\centering
	\input{figures/tikz_snr_vs_ber.tex}
	\caption{\ac{ber} vs \ac{snr} for retrained and non-retrained \acp{cnn} after $D_\mathrm{cd}$ changed from $\SI{17}{\CD}$ to $\num{2.06}$ and $\SI{2.42}{\CD}$ respectively.}
	\label{fig:snr_vs_ber}
\end{figure}
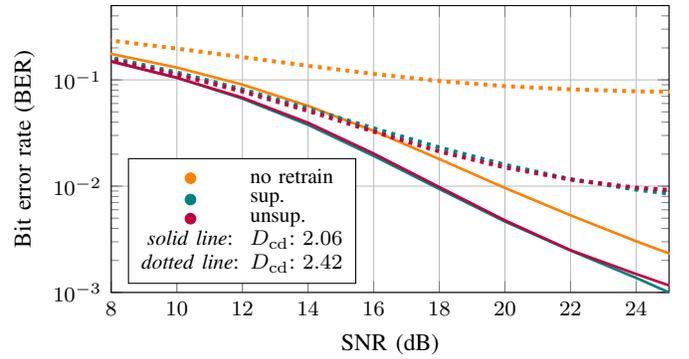

An additional comparison is shown in Fig. \ref{fig:snr_vs_ber}, where the performance of the non-retrained \ac{cnn} is compared with the models retrained for a $D_\mathrm{cd}$ of \SI{2.06}{\CD} and  \SI{2.42}{\CD} for different \acp{snr}. It can be seen that for a $D_\mathrm{cd}$ of \SI{2.06}{\CD}, the retrained models continuously outperform the non-retrained one by around \SI{2}{dB}. For a $D_\mathrm{cd}$ of \SI{2.42}{\CD}, the gap is even higher, as the non-retrained \ac{ber} flattens at around \num{8e-2}. For a \ac{ber} of \num{1e-1}, the gain of the retrained \acp{cnn} is around \SI{7}{dB}.

\subsection{Hardware Performance}
\label{subsec:hardware_performance}

In the following, we give the implementation results of our hardware architecture described in Sec. \ref{subsec:hardware_architecture} for \ac{pam}\nobreakdash-2. 
For \ac{fpga} implementation, \textit{Vivado HLS} in combination with \textit{Vivado Design Suite 2019.2} is used and the results are compared to the same \ac{cnn} running on two \acp{gpu}: the high-performance \ac{gpu} \textit{Nvidia RTX 2080} and the embedded \ac{gpu} \textit{Nvidia Xavier AGX}.
We implement our \ac{fpga} architecture on the \textit{ZCU102} evaluation board for a frequency of \SI{300}{\mega \hertz}. The power corresponds to the dynamic power given by Vivado Power Estimation Tool. For the \ac{gpu}, the dynamic power is obtained using \textit{nvidia\nobreakdash-smi}. The batch size of the \ac{gpu} implementations is increased until the \acp{gpu} run out of memory, while the \ac{dop} of the \ac{fpga} is adjusted to achieve maximal resource utilization.
The results are shown in Tab. \ref{tab:implementation_results}, where additionally to power and throughput, the time for retraining for a varying fiber dispersion factor, as discussed in \ref{subsec:hardware_performance} is given.

It is to note that a fair comparison to previous \ac{ann} \ac{fpga} implementations is not straightforward, as they are either based on a different topology, a different target platform or do not provide an implementation of the \ac{ann} training.  

\newcommand*{\myalign}[2]{\multicolumn{1}{#1}{#2}}

\begin{table}[t]

\caption{Hardware implementation results}
\label{tab:implementation_results}

\begin{tabular}{ccccccc}

\toprule
\multirow{2}{*}{Platform}                       & TP   & P   & Retraining    & LUT    & DSP               & BRAM  \\
& (\si{\mega bit}) & (\si{\watt}) & time (\si{\milli\second}) &  (\si{\percent}) & (\si{\percent}) & (\si{\percent}) \\

\midrule
\textit{ZCU102} &  \num{1200}         &  \num{4.83}     &  \num{3.3}        & \num{80.2}   & \num{69.4}     &  \num{15.8}       \\

\textit{RTX 2080} &  \num{140}                   &  \num{58}       & \num{29}         &  - & - & -                     \\
\textit{AGX Xavier} &  \num{11.8}                   &  \num{3.8}    & \num{340}            & - & - & -                  \\

\bottomrule
\end{tabular}

\end{table}


As compared to the \acp{gpu}, our \ac{fpga} architecture outperforms both implementations by orders of magnitude with respect to throughput and retraining time. Compared to the \textit{RTX 2080}, the \ac{fpga}'s throughput is \num{10} times higher, while the \textit{AGX Xavier} is outperformed by a factor of \num{100}. The dynamic power consumption of the \ac{fpga} architecture is slightly higher than that of the embedded \ac{gpu}, whereas it increases by a factor of \num{10} for the high-performance \ac{gpu}. 
One reason for the low throughput achieved by the \acp{gpu} is the small size of the \ac{ann}, which results in a high batch size required to fully utilize the \ac{gpu}. Thus, the memory bandwidth becomes the bottleneck of the \ac{gpu} implementations. 

\begin{figure}[b]
	\centering
	\input{figures/tikz_unrolling_analysis.tex}
	\caption{Dynamic power consumption and time to perform retraining for changing channel conditions on the \ac{fpga} for different \ac{dop}. The \ac{dop}, corresponding to the parallel calculated \ac{mac} operations in one layer, is given in the rectangle next to each point.}
	\label{fig:unrolling}
\end{figure}
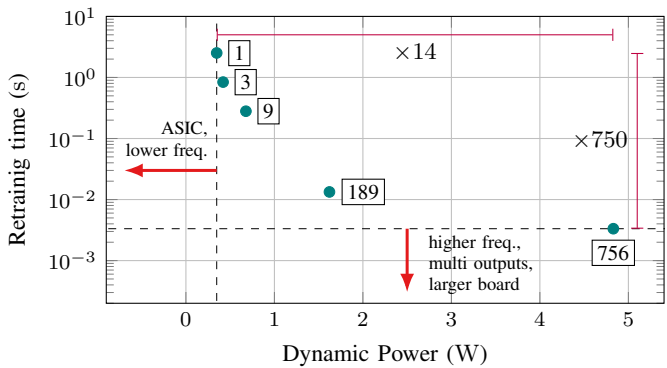

In Fig. \ref{fig:unrolling} we demonstrate the flexibility of our \ac{fpga} architecture. Each point corresponds to an implementation with a different \ac{dop} which can be loaded onto the \ac{fpga}. It can be seen that our architecture and \ac{fpga} as a platform allow to adapt to different application requirements regarding power consumption and retraining time. Those requirements could for example be imposed by a limited energy budget of the device or by the coherence time of the channel. 
In particular, the power consumption of \rectangled{1} is $14 \times$ lower than that of \rectangled{756}, while its retraining time is $750 \times$ higher. In between there exist multiple Pareto optimal points, which can also be loaded onto the same \ac{fpga}. Moreover, power consumption and retraining time could be further reduced as indicated by the red arrows.
It is important to highlight, that during retraining, the net datarate is only decreased for the supervised loss. For our novel unsupervised loss function, there is no downtime of the communication, as no labels need to be sent.

In summary, the results demonstrate that our architecture is adjustable to different application requirements like low power (order of \SI{100}{\milli \watt}) and high throughput (order of \si{\giga bit \per \second}). Our hardware implementation of a trainable ANN-based equalizer is considered a proof-of-concept and a first step towards practical systems in the field of optical communication. In future work, higher throughput could be achieved by designing an \ac{asic} and applying further optimizations, using a higher number of parallel outputs or more instances of the \ac{ann}.

%% file: figures/tikz_retraining_dispersion.tex
\begin{tikzpicture}[tight background,
	/pgfplots/every mark/.append style={solid, mark size=3.5pt},
	/pgfplots/every axis/.append style={font=\small}, 
	/pgfplots/every tick label/.append style={font=\footnotesize},
	/pgfplots/every axis legend/.append style={nodes={right},font={\footnotesize},row sep=-2pt}  
	]
	\begin{axis}[
            ymin=0.0001,
            ymax=2e-1,
            ymode=log, 
            xmin=17,
            xmax=26,
            restrict x to domain = 17:26,
            legend pos=south east,
            xlabel={$D_\mathrm{cd}$ ($\si{\CD}$)},
            ylabel={Bit error rate (BER)},
            width=\columnwidth,
            height=\columnwidth*0.6,
            xmajorgrids,
	    ymajorgrids
            ]



    \addplot [black, solid,line width=1pt,  mark=x, mark options={scale=1.2,solid}] table [x=fiber_dispersion_parameter,y=baseline]{data/Retraining/dispersion_new/8_to_26.txt}; \addlegendentry{$\textrm{CNN}_{\textrm{\scriptsize baseline}}$};
    
    \addplot [ALUColor4, solid, width=1pt,mark=*, mark options={scale=1.2,solid}] table [x=fiber_dispersion_parameter,y=no_retrain]{data/Retraining/dispersion_new/8_to_26.txt}; 	\addlegendentry{$\textrm{CNN}_{\textrm{\scriptsize no retrain}}$};

    \addplot [teal, solid, line width=1pt,mark=square*, mark options={scale=1.2,solid}] table [x=fiber_dispersion_parameter,y=sup_cont]{data/Retraining/dispersion_new/8_to_26.txt}; 	\addlegendentry{$\textrm{CNN}_{\textrm{\scriptsize sup.}}$};

    \addplot [purple , solid, line width=1pt,mark=diamond*, mark options={scale=1.2,solid}] table [x=fiber_dispersion_parameter,y=unsup_cont]{data/Retraining/dispersion_new/8_to_26.txt}; 	\addlegendentry{$\textrm{CNN}_{\textrm{\scriptsize unsup.}}$};

    \addplot [ALUColor2, solid,line width=1pt,mark=pentagon*, mark options={scale=1.2,solid}] table [x=fiber_dispersion_parameter,y=volterra]{data/Retraining/dispersion_new/8_to_26.txt}; 	\addlegendentry{$\textrm{Volterra}_{\textrm{\scriptsize sup.}}$};

	\end{axis}
\end{tikzpicture}

%% file: figures/tikz_retraining_dispersion_pam4.tex
\begin{tikzpicture}[tight background,
	/pgfplots/every mark/.append style={solid, mark size=3.5pt},
	/pgfplots/every axis/.append style={font=\small}, 
	/pgfplots/every tick label/.append style={font=\footnotesize},
	/pgfplots/every axis legend/.append style={nodes={right},font={\footnotesize},row sep=-2pt}  
	]
	\begin{axis}[
            ymin=0.01,
            ymax=2e-1,
            ymode=log, 
            xmin=17,
            xmax=26,
            restrict x to domain = 17:26,
            legend pos=south east,
            xlabel={$D_\mathrm{cd}$ ($\si{\CD}$)},
            ylabel={Bit error rate (BER)},
            width=\columnwidth,
            height=\columnwidth*0.6,
            xmajorgrids,
	    ymajorgrids
            ]



    \addplot [black, solid,line width=1pt,  mark=x, mark options={scale=1.2,solid}] table [x=fiber_dispersion_parameter,y=baseline]{data/Retraining/dispersion_new/8_to_26_pam4.txt}; \addlegendentry{$\textrm{CNN}_{\textrm{\scriptsize baseline}}$};
    
    \addplot [ALUColor4, solid, width=1pt,mark=*, mark options={scale=1.2,solid}] table [x=fiber_dispersion_parameter,y=no_retrain]{data/Retraining/dispersion_new/8_to_26_pam4.txt}; 	\addlegendentry{$\textrm{CNN}_{\textrm{\scriptsize no retrain}}$};

    \addplot [teal, solid, line width=1pt,mark=square*, mark options={scale=1.2,solid}] table [x=fiber_dispersion_parameter,y=sup_cont]{data/Retraining/dispersion_new/8_to_26_pam4.txt}; 	\addlegendentry{$\textrm{CNN}_{\textrm{\scriptsize sup.}}$};

    \addplot [purple , solid, line width=1pt,mark=diamond*, mark options={scale=1.2,solid}] table [x=fiber_dispersion_parameter,y=unsup_cont]{data/Retraining/dispersion_new/8_to_26_pam4.txt}; 	\addlegendentry{$\textrm{CNN}_{\textrm{\scriptsize unsup.}}$};

    \addplot [ALUColor2, solid,line width=1pt,mark=pentagon*, mark options={scale=1.2,solid}] table [x=fiber_dispersion_parameter,y=volterra]{data/Retraining/dispersion_new/8_to_26_pam4.txt}; 	\addlegendentry{$\textrm{Volterra}_{\textrm{\scriptsize sup.}}$};

	\end{axis}
\end{tikzpicture}

%% file: figures/tikz_snr_vs_ber.tex
\pgfplotsset{
    legend image with text/.style={
        legend image code/.code={%
            \node[anchor=center] at (0.3cm,0cm) {#1};
        }
    },
}

\begin{tikzpicture}[tight background,
	/pgfplots/every mark/.append style={solid, mark size=3.5pt},
	/pgfplots/every axis/.append style={font=\small}, 
	/pgfplots/every tick label/.append style={font=\footnotesize},
	/pgfplots/every axis legend/.append style={nodes={right},font={\footnotesize},row sep=-2pt}  
	]
	\begin{axis}[
            ymin=0.001,
            ymax=5e-1,
            ymode=log, 
            xmin=8,
            xmax=25,
            restrict x to domain = 0:28,
            legend pos=south west,
            xlabel={SNR (dB)},
            ylabel={Bit error rate (BER)},
            width=\columnwidth,
            height=\columnwidth*0.6,
            xmajorgrids,
	    ymajorgrids,
            legend cell align=left,
            ]


       \addlegendimage{only marks,ALUColor4}
    \addlegendentry{no retrain}

    \addlegendimage{only marks,teal}
    \addlegendentry{sup.}

    \addlegendimage{only marks, purple}
    \addlegendentry{unsup.}

    \addlegendimage{legend image with text=\textit{solid line}: }
    \addlegendentry{$D_\mathrm{cd}$: \num{2.06}}

    \addlegendimage{legend image with text=\textit{dotted line}: }
    \addlegendentry{$D_\mathrm{cd}$: \num{2.42}}




    \addplot [ALUColor4, solid,line width=1pt,  mark=none] table [x=SNR,y=no_retrain]{data/SNR_vs_BER/dispersion_2.06.txt}; 


   \addplot [teal, solid,line width=1pt,  mark=none] table [x=SNR,y=cont_supervised]{data/SNR_vs_BER/dispersion_2.06.txt}; 

   \addplot [purple, solid, line width=1pt,  mark=none] table [x=SNR,y=cont_unsupervised]{data/SNR_vs_BER/dispersion_2.06.txt}; 

    \addplot [ALUColor4, dotted,line width=1.5pt,  mark=none] table [x=SNR,y=no_retrain]{data/SNR_vs_BER/dispersion_2.42.txt}; 
    

   \addplot [teal, dotted,line width=1.5pt,  mark=none] table [x=SNR,y=cont_supervised]{data/SNR_vs_BER/dispersion_2.42.txt}; 

   \addplot [purple, dotted,line width=1.5pt,  mark=none] table [x=SNR,y=cont_unsupervised]{data/SNR_vs_BER/dispersion_2.42.txt}; 

	\end{axis}
\end{tikzpicture}

%% file: figures/tikz_unrolling_analysis.tex
\begin{tikzpicture}[
        square/.style={regular polygon,regular polygon sides=4},
        tight background,
	/pgfplots/every mark/.append style={solid, mark size=3.5pt},
	/pgfplots/every axis/.append style={font=\small}, 
	/pgfplots/every tick label/.append style={font=\footnotesize},
	/pgfplots/every axis legend/.append style={nodes={right},font={\footnotesize},row sep=-2pt}  
	]
	\begin{axis}[
            ymin=0.0002,
            ymax=10,
            ymode=log, 
            xmin=-0.9,
            xlabel={Dynamic Power (\si{\watt})},
            ylabel={Retrainig time (\si{\second})},
            width=\columnwidth,
            height=\columnwidth*0.6,
            xmajorgrids,
	    ymajorgrids
            ]

    \addplot [only marks, mark=*, mark options={teal, scale=1}] table [x=power,y=convergence_time]{data/unrolling_analysis.txt}; 
    \draw[line width=0.1mm, dashed] (-5, 0.00333) -- (10, 0.00333);
    \draw[line width=0.1mm, dashed] (0.348, 0.00002) -- (0.348, 10);

    \draw [-{Latex[length=2.5mm]}, ALUColor1, line width=0.4mm] (2.5, 0.00333) -- node [text=black, text width=1.5cm, midway, right=1.5mm, font=\scriptsize, yshift=-0.5mm] {higher freq., multi outputs, larger board} (2.5, 0.0003);

    \draw [-{Latex[length=2.5mm]}, ALUColor1, line width=0.4mm] (0.348, 0.03) -- node [text=black, text width=1.5cm, midway, above=0.1mm, font=\scriptsize, align=right, xshift=-2.5mm] {ASIC,\\lower freq.} (-0.7, 0.03);

    \draw[|-|, purple] (0.348, 5) -- node [midway, below] {\small \textcolor{black}{$\times 14$}}  (4.83, 5);   
    \draw[|-|, purple] (5.1, 2.52) -- node [midway, left] {\small \textcolor{black}{$\times 750$}}  (5.1, 0.00333);   

    \draw [black] (0.348, 2.52) node[xshift=3mm] {\footnotesize \rectangled{1}};
    \draw [black] (0.42, 0.84) node[xshift=3mm] {\footnotesize \rectangled{3}};
    \draw [black] (0.678, 0.28) node[xshift=3mm] {\footnotesize \rectangled{9}};
    \draw [black] (1.622, 0.01333) node[xshift=4.5mm] {\footnotesize \rectangled{189}};
    \draw [black] (4.83, 0.00333) node[yshift=-3.2mm] {\footnotesize \rectangled{756}};

    \end{axis}


\end{tikzpicture}

%% file: chapters/conclusion.tex
\section{Conclusion}
\label{Sec:conclusion}

In this work, we propose a novel approach for unsupervised retraining of an \ac{ann}-based equalizer for changing channel conditions. Therefore, we present an unsupervised loss function for \ac{pam}-2 and \ac{pam}-4 modulation and demonstrate its ability to adapt to a varying fiber dispersion parameter. Furthermore, we present a pipelined \ac{fpga} architecture of our approach, to bridge the gap between \ac{ann}-based communication algorithms and efficient hardware implementation.
As a result, we demonstrated that our unsupervised approach nearly reaches the communication performance of supervised retraining, while reducing the overhead of pilot symbols as labels. Moreover, we show that a throughput in the order of \si{\giga bit \per \second} is feasible with our \ac{fpga} implementation, which can't be achieved by a high-end \acp{gpu}, while it is also highly flexible. 
